\documentstyle[preprint,pre,aps]{revtex}
\begin{document}

\title{%
Eigenvalue spectrum of the Frobenius-Perron operator
near intermittency}
\author{Z.~Kaufmann}
\address{Institute for Solid State Physics,
E\"otv\"os University,
P. O. Box 327, H-1445 Budapest, Hungary}
\author{H.~Lustfeld}
\address{Institut f\"ur Festk\"orperforschung,
Forschungszentrum J\"ulich,
D-52425 J\"ulich, Germany}
\author{J.Bene}
\address{Institute for Solid State Physics,
E\"otv\"os University,
P. O. Box 327, H-1445 Budapest, Hungary}
\date{\today}
\maketitle

\mediumtext
\begin{abstract}%
The spectral properties of the Frobenius-Perron operator
of one-dimensional maps are studied when approaching a weakly
intermittent situation.
Numerical investigation of a particular family of maps shows that the
spectrum becomes extremely dense and the eigenfunctions become
concentrated in the vicinity of the intermittent fixed point.
Analytical considerations generalize the results to a broader class of
maps near and at weak intermittency and show that one branch of the
map is dominant in determination of the spectrum.
Explicit approximate expressions are derived for both the eigenvalues
and the eigenfunctions and are compared with the numerical results.
\end{abstract}
\pacs{05.45.+b}
\narrowtext

\section{%
Introduction}

Correlation functions play an important role
in the characterization of chaotic systems.
In certain classes of systems they decay exponentially.
Cases when it is proven are Axiom A
systems~\cite{Bb,P,Rl}, mixing maps~\cite{Rlb} and maps which
are analytic on a set of rectangles~\cite{R}.
It is generally believed that hyperbolic systems usually exhibit an
exponential correlation decay, although the rigorous proof of this
can be quite a difficult task in particular cases ~\cite{Li}.
On the other hand there are similarly
important classes of systems where the decay of correlations
is slower than exponential. Such a behavior has been found
in intermittent one-dimensional maps~\cite{GH} and in Hamiltonian
systems with mixed phase space~\cite{GK}.
Intermittency, the alternation of almost regular and chaotic
motion, can precede birth of a pair of stable-unstable fixed point or
periodic orbit (tangent bifurcation)~\cite{MP,PM},
can be caused by a marginally unstable fixed
point~\cite{M,GH,SzGy,SzTCsK,OM}
or the marginal behavior at the boundary of an island in general
Hamiltonian systems~\cite{GK}.
It can also arise when a chaotic attractor becomes unstable due
to crises~\cite{GOY},
effect of noise or driving of another attractor~\cite{PSTr,CL}.
Intermittency is known to possess several special properties.
The spectrum of R\'enyi-entropies shows a phase-transition-like
behavior~\cite{SzTCsK,CsSz,BSzF,KSz,SzTV},
the decay properties of truncated entropies,
complexity and the dynamical fluctuations
also differ from those of expanding maps~\cite{SzGy,KSz,Sz8,Sz9,GW,W}.
The kind of intermittency we study is due to a marginally unstable
fixed point \cite{M,GH,SzGy,SzTCsK,OM}.
The vicinity of the fixed point is visited by typical trajectories
and the time needed to leave this vicinity scales with a power
of the distance from the fixed point instead of the logarithm as for
an unstable fixed point.
Depending on the actual map there can even
exist a smooth invariant density
to which initial densities converge, as in case of our example.
This case is called weak intermittency, otherwise we talk about strong
intermittency~\cite{GH}.

The decay of correlations may be related to the spectral properties of
the Frobenius-Perron operator describing the time evolution of the
probability density given in the phase space of the system in
question.
The rough
picture is that the largest eigenvalue is unity (except for transient
chaos~\cite{T2}, not to be treated here)
and the corresponding eigenfunction is related to the
natural measure, while the leading asymptotics of the correlation
decay is related to a spectral gap
between the largest eigenvalue and the rest of the
spectrum. Note that in two or higher dimensional systems
even in hyperbolic cases
the direct study of the spectrum of the Frobenius-Perron operator
involves a series of difficulties, related to both the suitable
definition of the Frobenius-Perron operator and the relevant function
space in which the eigenvalue problem is to be solved.

In case of one-dimensional maps the Frobenius-Perron operator
takes the form
$L\varphi(x)=\sum_{z:f(z)=x}\frac{\varphi(z)}{|f'(z)|}$.
Here the sum goes over the preimages $z$ of $x$.
Existence of a unique invariant density (belonging to eigenvalue
$\lambda_0=1$) is proven for expanding
piecewise ${\rm C}^2$ maps~\cite{LY} and fully developed maps with
negative Schwarzian derivative~\cite{Mi}.
If a unique ergodic measure exists the correlation function can be
expressed
in terms of the Frobenius-Perron operator.
If the spectrum is discrete
eigenvalues $\lambda_n$ and eigenfunctions $\varphi_n$
may be used to explore the time evolution of an
initial probability density $\varphi(0,x)$.
Then decay of correlations and
the convergence of $\varphi(t,x)$ towards the
invariant density is governed by $\lambda_1$
(where $\lambda_0>|\lambda_1|\ge|\lambda_2|
\ge\ldots\ge|\lambda_n|\ge\ldots$).

The relation between the spectral properties
and the correlation decay becomes much more complex
if the latter one is slower than exponential.
In such a case, arising typically in intermittent systems,
the largest eigenvalue must be an accumulation point of the spectrum.
The standard methods~\cite{SzT,T,Rl,C,CPR} developed
for determining the eigenvalues of operators similar to
the Frobenius-Perron operator
can not be used to determine the non-leading
eigenvalues when the system is close to an intermittent situation.
A direct study of the spectrum of generalized
Frobenius-Perron operators in case of intermittency can be found
in Ref.~\cite{Pr}.
Working in the space of functions with bounded variation
a nonzero essential spectral radius~\cite{Rlc} is obtained,
which reaches $\lambda_0$ for the case of $L$~\cite{Pr}.
Study of the spectrum for intermittency that
precedes tangent bifurcation is presented in~\cite{BaERl}.

Our motivation has been to calculate the spectral properties near
and at
intermittency and to establish their link to the correlation decay.
In the present paper we concentrate on the first task,
while the derivation of the correlation functions
will be published elsewhere~\cite{LBK}. In order to avoid
the difficulties arising in higher dimensional systems
we consider piecewise analytic one-dimensional
maps and choose the function space of real analytic functions.
Note that other methods and other function
space has been used than in Ref.~\cite{Pr},
which allows us a more detailed description.
The paper is organized as follows.
In Section 2 we study the spectral properties of the Frobenius-Perron
operator numerically in a family of fully developed chaotic
maps~\cite{GySzJ,GySzZ},
where the parameter is tuned up to weak intermittency.
Some aspects of the numerical work are discussed in Appendix A.
Section 3 contains the analytical results, which are valid for
a wider class of maps.
Details of the calculation are given in Appendix B.

\section{%
A model and numerical calculation}

Our model is the piecewise parabolic map~\cite{GySzZ}.
This is probably the simplest
dynamical system which can show intermittency.
The general form of the map is
\begin{equation}
f_r(x)=\frac{1+r-\sqrt{(1-r)^2+4r|1-2x|}}{2r}\;.\label{fpp}
\end{equation}
It maps the interval $[0,1]$ twice onto itself for every
value of the parameter $r$ in the interval $[-1,1]$\,. Its
name comes from the fact that its inverse has parabolic
branches. The point $x=0$ is a fixed point which becomes
marginal in the intermittent case $r=1$~\cite{GySzZ,Hem,GH}.
For the other special case $r=0$ we obtain the uniformly
expanding tent map.
A special property of the piecewise parabolic map is that the
stationary probability density has a simple form~\cite{GySzZ},
$p_r(x)=1+r(1-2x)$.
This is still normalizable for $r=1$,
which means that the map possesses weak intermittency.
Note that the average Ljapunov exponent is positive.

In order to study the eigenvalue spectrum of the
Frobenius-Perron operator we derive a matrix representation on
the basis of the Legendre polynomials $P_k$.
After normalization $Q_k(x)=\sqrt{k+1/2}P_k(x)$
these polynomials form a complete orthonormal
system on the interval $[-1,1]$\,.
In order to use them we have to transform the
map and this basis to the same interval.

In the case of our model the action of
the Frobenius-Perron operator on an arbitrary function
$\varphi(x)$ can be given by
\begin{equation}
L\varphi(x)=\left[\varphi\left(F_1(x)\right)+
\varphi\left(F_2(x)\right)\right]
\left|F_2'(x)\right|\;,\label{Lfi}
\end{equation}
where
$
F_{1,2}(x)=\frac12 \mp\left(\frac12-\frac{1+r}2 x+\frac r2 x^2\right)
$
stands for the inverse branches of the map.

Expanding the function $\varphi(x)$ in terms of $Q_k(x)$ as
$\varphi(x)\approx\sum_{k=0}^\infty b_kQ_k(x)$
we obtain the (infinite) matrix
representation of the Frobenius-Perron operator,
\begin{equation}
L\varphi(x)=\sum_{k,l=0}^\infty L_{lk}b_k Q_l(x)\;,
\end{equation}
where the matrix elements $L_{lk}$ are of the form
$L_{lk}=\int_{-1}^1 Q_l(x) L Q_k(x)\,dx$\,.
In the numerical calculation we have truncated this matrix confining
the indices between $0$ and $N$.
The integrals have been evaluated by numerical integration.
Exploiting the symmetry of the map the matrix size has been
effectively reduced by a factor of 2.
Then the eigenvalue spectrum
$\lambda_n^{(N)},\; n=0,1,2,\ldots$ of the matrix has been
determined. It was quite remarkable that the eigenvalues turned out to
be always positive real numbers.
This calculation was
performed for different values of the parameters $r$ and $N$.
Fig.\ 1 shows the spectrum as a function of the parameter $r$ at
two different values of $N$.
The solid parts of the curves in Fig.\ 1 correspond to cases
where, at the given value of $N$, the precision of the eigenvalues
have reached $0.003$ (which is approximately the
resolution of the figure).
The precision was determined by comparing results for
different values of $N$.

It is seen that all the eigenvalues $\lambda_n$ with fixed $n$ tend to
value $1$ for $r\rightarrow 1$.
On the other hand we find that near any fixed $\lambda$ value
the density
of the eigenvalues increases suggesting a continuous spectrum at
intermittency.
The eigenfunctions of the operator also show a remarkable behavior
when approaching the intermittent case $r=1$. Apart from the first
one, they become more and more concentrated in the vicinity of the
fixed point (see Fig.\ 2). That means, the limiting eigenfunctions
are singular.
One might expect the strange case that the eigenvalue $\lambda_0=1$
becomes infinitely degenerate in the space of analytic functions.
This is avoided since the eigenfunctions become singular in the limit.
At the same time the tails of the eigenfunctions outside the vicinity
of the fixed point are smooth functions and become more and more
similar to the invariant density $p(x)$ (see inset in Fig.\ 2).
In fact, with a suitable normalization they converge to $p(x)$.

It has been observed that for
$r\le 0.9$ the eigenvalues converge quite fast when increasing
the value of $N$. Going closer to the case $r=1$ the
convergence becomes slower and the limiting eigenvalues get closer to
the largest one ($\lambda_0=1$). In the case $r=1$ the
eigenvalues converge to unity, but slower than exponentially.
At the same time the eigenvalues with higher index converge
slower, consequently
the closer $r$ is to unity the less eigenvalues are precise.
Its reason is twofold. On the one hand, the spectrum becomes
denser and denser when approaching the intermittent case, which
certainly leads to numerical problems. On the other hand, the
eigenfunctions $\varphi_n(x)$ close to the intermittent situation after
suitable conjugation can be well approximated by powers $x^{n-1}$. As
it is explained in Appendix A this leads to the problem, that only a
few eigenvalues and eigenfunctions can be obtained numerically.

\section{%
Analytical results}

We can understand the behavior of the spectrum near the intermittent
case by keeping only the term of the Frobenius-Perron operator
$L\varphi(x)=\sum_{z:f(z)=x}\frac{\varphi(z)}{|f'(z)|}$
that corresponds to the lower inverse branch.
This branch contains the intermittent fixed point.
(We shall denote the corresponding operator by $L_1$, while the
operator containing the upper inverse branch of the map will be
denoted by $L_2$.)
The first explanation is the following.
The numerical results show that the
eigenfunctions, except for the invariant density $p(x)$, are
concentrated in the vicinity of the fixed point. Therefore the most
important region is this neighbourhood.
The contribution of $L_2$ comes
from the rightmost piece of the interval [0,1],
while this piece is reached from the neighbourhood of the maximum
in one iteration. At the same time the eigenfunctions are smooth and
(except for $p(x)$) have quite low value there compared to the
vicinity of the fixed point.
Therefore the contribution of the upper branch is
small and is a slowly varying function.
So we first study $L_1$ and later we return to the full
Frobenius-Perron operator to estimate the effect of the upper branch.
The considerations are quite general, we only assume that
the map is single-humped,
we assume special properties of the map at the endpoints and the
maximum,
and assume that the map is smooth and expanding everywhere else.

Without restricting generality we let the map act on the interval
$[0,1]$ so that the fixed point is at $x=0$. We
assume that the map satisfies the following formulas
such as model (\ref{fpp}) does:
\begin{eqnarray}
f(x)&=&(1+\varepsilon)x+a x^2+{\cal O}(x^3)\;\;\;
\mbox{if}\;\; x\ll 1 \;, \label{f}\\
f(x)&=&d(1-x) +{\cal O}((1-x)^2)\;\;\; \mbox{if}\;\; 1-x\ll 1 \;,\\
F_{1,2}(x)&=&\hat x \mp (g_{1,2}\varepsilon(1-x)+h_{1,2}(1-x)^2)
\nonumber\\
&&\hspace{12.6mm}\mbox{}+{\cal O}((1-x)^3)\;\;\;
\mbox{if}\;\; 1-x\ll 1\;,\label{fc}
\end{eqnarray}
where $\hat x$ is the location of the maximum, $\varepsilon, a, d,
g_{1,2}, h_{1,2}$ are suitable positive constants
and $F_{1,2}$ are the inverse branches of $f(x)$.
For $\varepsilon\rightarrow 0$ keeping $a, d, g_{1,2}, h_{1,2}$ fixed
we are approaching the intermittent situation.
The eigenvalue equation for $L_1$ takes the form
\begin{equation}
\lambda\varphi(f(x))=\frac{\varphi(x)}{|f'(x)|}\;,\;\;
0\le x\le\hat x\;.\label{L1}
\end{equation}
We can solve it in the neighbourhood of the origin by introducing
$\omega(x)=\ln\varphi(x)$\,.
Since $f(x)-x$ is small when we are close to the intermittent case
$\omega(f(x))-\omega(x)$
can be approximated by the linear term of Taylor expansion.
Then one obtains
\begin{equation}
\omega'(x)=-\frac{\ln f'(x)+\ln\lambda}{f(x)-x}+
{\cal O}(\varepsilon^2)
\end{equation}
Substitution of (\ref{f}) and integration yields
\begin{equation}
\varphi(x)\approx
c\frac{x^{\nu(\lambda)}}{(x+\frac{\varepsilon}{a})^{\nu(\lambda)+2}}
\;,\label{finu}
\end{equation}
where
$\nu(\lambda)=\frac{-\ln\lambda}{\varepsilon}-1+\frac{\varepsilon}{2}$.
Let us restrict ourselves to the case of analytic eigenfunctions,
that means $\nu(\lambda)=n-1$, where $n$ is a positive integer.
Then
\begin{equation}
\varphi_n(x)\approx
c\frac{x^{n-1}}{(x+\frac{\varepsilon}{a})^{n+1}}\;.\label{fi}
\end{equation}
The corresponding eigenvalue is obtained from
the expression below (\ref{finu}) as
$\lambda_n\approx e^{-n\varepsilon}$.

It is interesting to see that even the exact eigenvalues of $L_1$
(see(\ref{L1})) for a not necessarily small $\varepsilon$
can be obtained from the stability of the fixed point.
The reason is the following.
If we know the eigenfunction $\varphi(x)$ in any small neighbourhood
$V_0$ of the
fixed point, then by (\ref{L1}) we can determine it in the image
$V_1$ of this
interval. Repeating this procedure, say $k$ times,
the part of the interval
$V_k$ lying outside the previous one $V_{k-1}$
will be mapped outside $V_k$ .
Hence we can find the eigenfunction satisfying (\ref{L1}) in
$[0,\hat x]$. For small enough neighbourhood $V_0$
the function $f(x)$ can be approximated linearly, which yields
eigenfunctions $x^{n-1}$ and eigenvalues
\begin{equation}
\lambda_n=(1+\varepsilon)^{-n}\;,\;\;n=1,2,\ldots\;. \label{lae}
\end{equation}
Moreover, these are the exact eigenvalues of $L_1$, since the
relative precision of the linear approximation
can be arbitrarily increased by shrinking the
interval $V_0$. However, for the determination of the eigenfunctions
the quadratic term in (\ref{f}) is important.

After these considerations we can estimate the effect of the
upper inverse branch of the map.
The eigenvalue equation may be written as
\begin{eqnarray}
\lambda f_1'(x)\varphi(f(x))&=&\varphi(x)+\chi(x)\;,\label{L2}\\
\chi(x)&=&\varphi(F_2(f(x)))f_1'(x)F_2'(f(x))\;.\label{psi}
\end{eqnarray}
It is seen from Eq.\ (\ref{fi}) that very close to the fixed point the
eigenfunction starts as
$\varphi_n(x)\approx\beta x^{n-1}$
(we do not fix the normalization here, $\beta$ may contain a power
of $\varepsilon$)
and it is concentrated in a region of width of order $\varepsilon$.
We assume that the eigenfunction gets small
corrections compared to Eq.\ (\ref{fi}).
After estimating the value of $\varphi_n$ and writing down power-series
forms for it in the neighbourhood of $x=0$ and $\hat x$
we obtain results for
$\lambda_n$ and coefficients of the expansion
$\varphi_n(x)=\sum_{k=0}^{\infty} \beta_k x^k$\,:
\begin{eqnarray}
\beta_m&=&{\cal O}(\beta\varepsilon^{n-m})\;,\;\;\mbox{if}\;\;m<n-1\;,
\label{bem}\\
\lambda_n&=&(1+\varepsilon)^{-n}+{\cal O}(\varepsilon^2)\;.\label{lak}
\end{eqnarray}
The actual value of the correction depends on the shape of the map but
its order of magnitude does not. The order
of the correction in (\ref{lak})
was also shown by the numerical results for the map
(\ref{fpp}) for the $n=1,2,3$ eigenvalues.
For the perturbation of the eigenfunction (see (\ref{bem}))
the case $m=0$ shows that the eigenfunction is shifted, and the ratio
of this shift $\beta_0$ and the maximum is of order $\varepsilon$.
This was
numerically verified for $n=2,3$ (see the nonzero starting value in
Fig.\ 3).
The value $\beta_1$ gives the initial slope of the eigenfunction at the
fixed point. $\beta_1$ can be compared to the slope
${\cal O}(\beta\varepsilon^{n-2})$ of the straight line connecting
the origin
and the maximum of $\varphi_n$. Their ratio is also
proportional to $\varepsilon$,
what was numerically checked for $n=3$.

It is important to mention, that not only values of $\lambda_n,\beta_0$
and $\beta_1$ show good agreement between the analytic and numeric
results. The shape of the eigenfunction itself agrees very well,
too.
The agreement is surprisingly good for $r=0.999$\,.
Fig.\ 3 also supports
the above estimations, namely, that the deviation, which is
due to the effect of the upper branch, is proportional to $\varepsilon$.
One should notice, that we did not get the eigenvalue $\lambda_0=1$
this way. The reason is, that for $n=0$ the eigenfunction is
not concentrated in the vicinity of the fixed point, as indicated
in the beginning of this section.
This means, the contribution $\chi(x)$ of the upper inverse branch of
the map is not negligible in this case.

In the above estimation of the effect of the second branch $n$ was
assumed to be constant. However, the effect of $L_2$ remains also
small in the limit $r\rightarrow 1$ when  $n$ tends to infinity in
such a way that $\lambda_n$ is kept constant.
This can be realized using a series of $r_n$ values which satisfy
$\lambda_n(r_n)=\lambda$.
The effect of $L_2$, i.e.\ the difference $\delta\lambda_n$ of analytic
eigenvalues (\ref{lae}) for $L_1$ and numeric eigenvalues for $L$
was determined for several values of $n$ and $r$.
Since the spacing $\Delta\lambda_n=\lambda_n-\lambda_{n+1}$
of the eigenvalues
decreases in the limit the relative precision
$\delta\lambda_n/\Delta\lambda_n$ is important.
Fig.\ 4 shows that this ratio is roughly constant in function of $n$
(depends on $\lambda$ only) when it is considered as a
function of $n$ and $\lambda$, and it is small for $\lambda\approx 1$.
According to these numerical findings one may assume that also the
expression (\ref{fi}) for the eigenfunctions is valid with a fixed
precision in the limit $\varepsilon\rightarrow 0$ for fixed $\lambda$.
Therefore an approximating eigenfunction can be obtained for the
intermittent case $r=1$ as the limit of (\ref{fi}),
which can also be calculated using (\ref{finu}),
the expression of $\nu(\lambda)$ and keeping
$\lambda$ fixed. This way one obtains
\begin{equation}
\varphi_\lambda(x)\approx
A_\lambda\frac 1{x^2} e^{-\frac{|\log\lambda|}{ax}+2ax}\;,
\end{equation}
while the spectrum is found to be continuous.

\section{Acknowledgement}

Authors thank T. T\'el for reading the manuscript and making
several valuable remarks.
This work has been supported by the International Relations Offices of
Germany and Hungary, project No X231.3 and by the Hungarian
Academy of Sciences under Grant Nos.\ OTKA 2090 and OTKA F.4286.
Two of the authors (J.B. and Z.K.)  are grateful for the hospitality
of the Institut f\"ur Festk\"orperforschung, Forschungszentrum
J\"ulich GmbH, where part of the work has been done.
One of the authors (H.L.)  is grateful for the hospitality of the
Institute for Solid State Physics, E\"otv\"os University Budapest,
where another part of the work has been done.

\begin{appendix}
\section{Representing the eigenfunctions on an orthonormal basis}

In this appendix we explain why an orthonormal basis is not well
suited to get many eigenvalues correctly near an intermittent
state.
This is easier to see after applying the following conjugation to
the piecewise parabolic map:
\begin{equation}
y=h(x)=\frac{x}{x+\frac{\varepsilon}{a}}
\left(1+\frac{\varepsilon}{a}\right)\;.
\end{equation}
Then the eigenfunctions of the map
can be well approximated by powers $\{y^n\}$ near intermittency.
Therefore $\langle\Phi_m|L|y^n\rangle$
is a good matrix representation of $L$,
where $\{\Phi_n\}$ is the adjoint basis to $\{y^n\}$.

Using an orthonormal basis $\{q_n(y)\}$ the matrix elements
$L_{nm} =\langle q_n|L|q_m\rangle$
can be obtained by the following transformation:
\begin{equation}
\langle\Phi_m|L|y^n\rangle=\sum_{l,k} W^{-1}_{mk}L_{kl}W_{ln}\;,
\end{equation}
where
$W_{nm}=\int_0^1 q_n(y)y^m\,dy$\,.
How many eigenvalues we can get correctly depends on how weakly
conditioned the matrix $W_{nm}$ is.

In our case ($q_n(x)$ is the normalized Legendre polynomial
transformed to $[0,1]$) the matrix elements can be calculated
analytically
using Rodrigues formula and applying partial integration:
\begin{eqnarray}
W_{nm}=
\left\{\begin{array}{ll}
{\displaystyle\frac{\sqrt{2n+1}(m!)^2}{(m-n)!(m+n+1)!}}
&\mbox{ for }n\leq m\\
0 &\mbox{ for }n>m
\end{array}\right.
\end{eqnarray}
The maximum number $N_{max}$ of eigenvalues
and eigenvectors that can be reliably computed
can be approximately read off
from the singular value decomposition representing a matrix $W$ as
$W_{i,j} = U_{il} D_{ll} V^T_{l,j}$.
Here $U$ and $V$ are orthogonal $n\times n$ matrices
and $D$ is diagonal one with the property
$D_{ii} \geq D_{jj} \geq 0 \mbox{ for } i<j$.
We have determined $N_{max}$ from the condition
\begin{equation}
D_{N_{max},N_{max}} >\delta,\;\;
D_{N_{max}+1,N_{max}+1} \le\delta
\end{equation}
with two different choices for $\delta$ ($10^{-5}$ and $10^{-10}$).
It is seen in Table 1, that $N_{max}$ increases slowly
with increasing $n$.
Therefore we conclude that close to intermittency
the shifted Legendre polynomials represent
a good basis to determine the first few eigenvalues and
eigenfunctions. Beyond that they are expected to produce
misleading results.
We surmise that these conclusions are intrinsic
for any orthogonal basis since the matrix $W$ is
nearly degenerate.
This corresponds to the fact that the basis functions $y^n$ are nearly
parallel for large $n$ when angles are defined through the usual
${\cal L}_2$ norm.

\section{Effect of the upper inverse branch}

To obtain estimations (\ref{bem}), (\ref{lak})
we start from Equations (\ref{L2})-(\ref{psi}). It follows, that
the order of the maximum value of $\varphi(x)$ is
$\beta\varepsilon^{n-1}$.
The numerical calculation not only supports these estimates but shows
that the function outside that region is smooth and slowly varying.
The integral of the eigenfunction should be zero since it
should be orthogonal to the first left eigenfunction of $L$. At the
same time its integral in the above mentioned region near 0
is of order $\beta\varepsilon^n$, hence
it should be of the same order
around the position $\hat x$ where $f$ has its maximum.
Thus the Taylor expansion
$\varphi(x)=\sum_{k=0}^{\infty}\hat c_k (x-\hat x)^k$
possesses coefficients of order $\beta\varepsilon^n$ at most,
$\hat c_k={\cal O}(\beta\varepsilon^n)$.
Going to the iterate of $\hat x$ we apply equation (\ref{L2}) for
$x\approx f(\hat x)=1$. Using Eq.\ (\ref{fc})
we obtain
estimation for the Taylor series around the point $x=1$.
The second term of (\ref{L2}) connects it to $\varphi(x)$ around
$x=0$\,. Since the expressions $F_2(f(x))$ and $f(x)$
in (\ref{psi}) start linearly for $x\approx 0$ the coefficients
in the expansions
$\chi(x)=\sum_{k=0}^{\infty} c_k x^k$ and
$\varphi(x)=\sum_{k=0}^{\infty} \tilde c_k (1-x)^k$
are of the same order of magnitude:
$c_0,\tilde c_0={\cal O}(\beta\varepsilon^{n+1})$,
$c_k,\tilde c_k={\cal O}(\beta\varepsilon^n)\;,\;\;k>0$.

Returning to the neighbourhood of the intermittent fixed point we
apply (\ref{L2}) for $x\approx 0$. We describe the
corrected eigenfunction by the expansion
$\varphi(x)=\sum_{k=0}^{\infty} \beta_k x^k$,
where $\beta_k,\; k<n-1$ are small and $\beta_{n-1}\approx\beta$,
which ensures that the eigenfunction gets small correction.
Substituting the expansions for $\varphi(x), \chi(x)$
and (\ref{f}) into (\ref{L2}),
expanding the powers and
collecting terms proportional to $x^m$ we obtain
\begin{eqnarray}
\lambda\sum_{k=Int(\frac{m}{2})}^{m-1}
b_{m,k}\alpha^{m-k}(1+\varepsilon)^{2k+1-m}\beta_k \hspace{15mm}
\nonumber\\
+\left[\lambda(1+\varepsilon)^{m+1}-1\right]\beta_m
=c_m+\sum_{k=0}^{m-2}{\cal O}(\beta_k)\;,\label{labe}
\end{eqnarray}
where
$b_{m,k}=\left({k\choose m-k}+2{k\choose m-k-1}\right)$,
and ${p\choose q}=0$ if $q>p$ or $q<0$\,.
Using this equation and the estimations for $c_k$ it can be shown
recursively, that
$\beta_m={\cal O}(\beta\varepsilon^{n-m})\;,\;\;\mbox{if}\;\;m<n-1$\,.
For $m=0$ only the term  containing $\beta_m$ remains on the
left hand side.
Since the eigenvalue (\ref{lae}) is expected to get small correction
the coefficient of $\beta_m$ in (\ref{labe})
is of order $\varepsilon$\,. This
yields $\beta_0={\cal O}(\beta\varepsilon^n)$.
For $m>0$ the sum in (\ref{labe}) is not suppressed by $c_m$,
so we get $\beta_m={\cal O}(\varepsilon^{-1}\beta_{m-1})$, which proves
our statement for $\beta_m$.
However, for $m=n-1$ the eigenvalue (\ref{lae}) would give zero
for the coefficient of $\beta_m$, therefore this coefficient depends
strongly on the perturbation of $\lambda$\,. At the same time in
the case $m=n-1$ the order of $\beta_m=\beta_{n-1}\equiv\beta$ is known,
hence this case determines
the corrected value of $\lambda$ with the help of $\beta_m$\,:
$\lambda_n=(1+\varepsilon)^{-n}+{\cal O}(\varepsilon^2)$.

\end{appendix}

\begin{figure}
\caption{%
Eigenvalues of the Frobenius-Perron operator $L$ for the piecewise
parabolic map with maximal matrix index
(a) $N=40$, (b) $N=80$.
Solid lines: precision higher than $0.003$;
dashed lines: precision lower than $0.003$.}
\end{figure}

\begin{figure}
\caption{%
The eigenfunction $\varphi_2$ for $r=0.9$ (solid line),
$r=0.97$ (dashed line) and $r=0.99$ (dotted line).
Inset:
The eigenfunctions $\varphi_0$ (solid line),
$\varphi_1$ (long dashes), $\varphi_2$ (short dashes),
$\varphi_3$ (dotted line) and $\varphi_4$ (dashed-dotted line),
for $r=0.999$ with different normalization.}
\end{figure}

\begin{figure}
\caption{%
Comparison of
eigenfunctions $\varphi_1,\varphi_2,\varphi_3$ and $\varphi_4$
of $L_1$ obtained analytically (solid lines) with corresponding
eigenfunctions of $L$ calculated numerically
for $r=0.99$ (dotted lines) and for $r=0.999$ (dashed lines)
in the vicinity of the fixed point.
For identification of the curves we note that
$\varphi_1<\varphi_2<\varphi_3<\varphi_4$
at the right edge of the figure.}
\end{figure}

\begin{figure}
\caption{%
Effect of the second branch on the eigenvalues compared to the
eigenvalue spacing in function of $\lambda$ and $n$.
$\Diamond: n=1,\; +: n=2,\; \Box: n=3,\; \times: n=4,\;
\triangle: n=5,\; *: n=6$.}
\end{figure}

\begin{table}[b]
\begin{center}
\begin{tabular}{|r|r|r|}
\hline
 $n$ & $\delta=10^{-5}$ & $\delta=10^{-10}$ \\
\hline
   5 &  5 &  5 \\
  10 &  8 & 10 \\
  20 & 10 & 16 \\
  50 & 12 & 21 \\
 100 & 14 & 24 \\
 500 & 18 & 32 \\
1000 & 20 & 35 \\
\hline
\end{tabular}
\end{center}
\caption{Values of $N_{max}$ in function of $n$
for different $\delta$.}
\end{table}

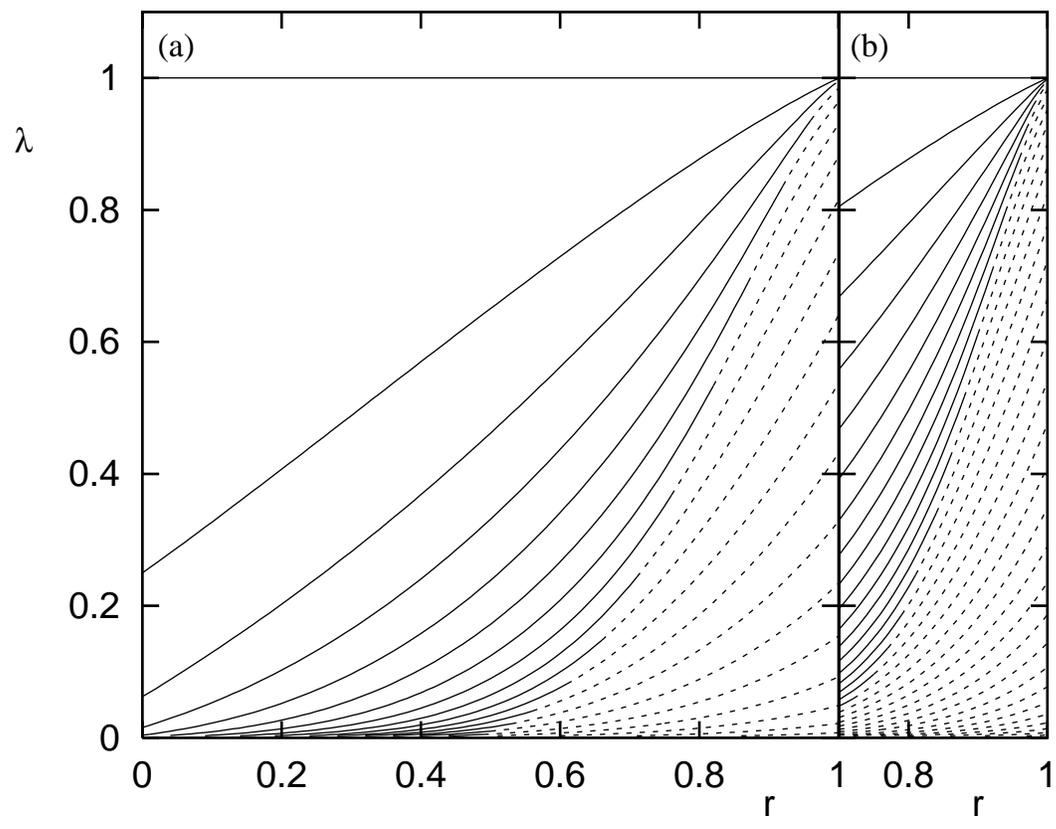
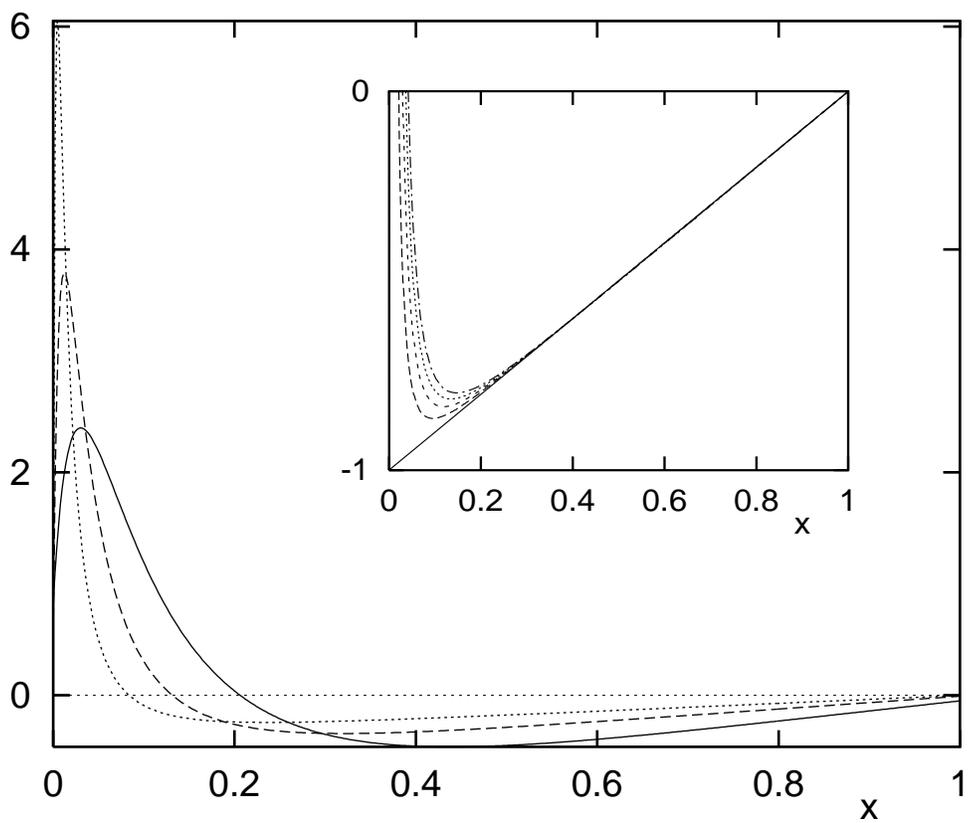
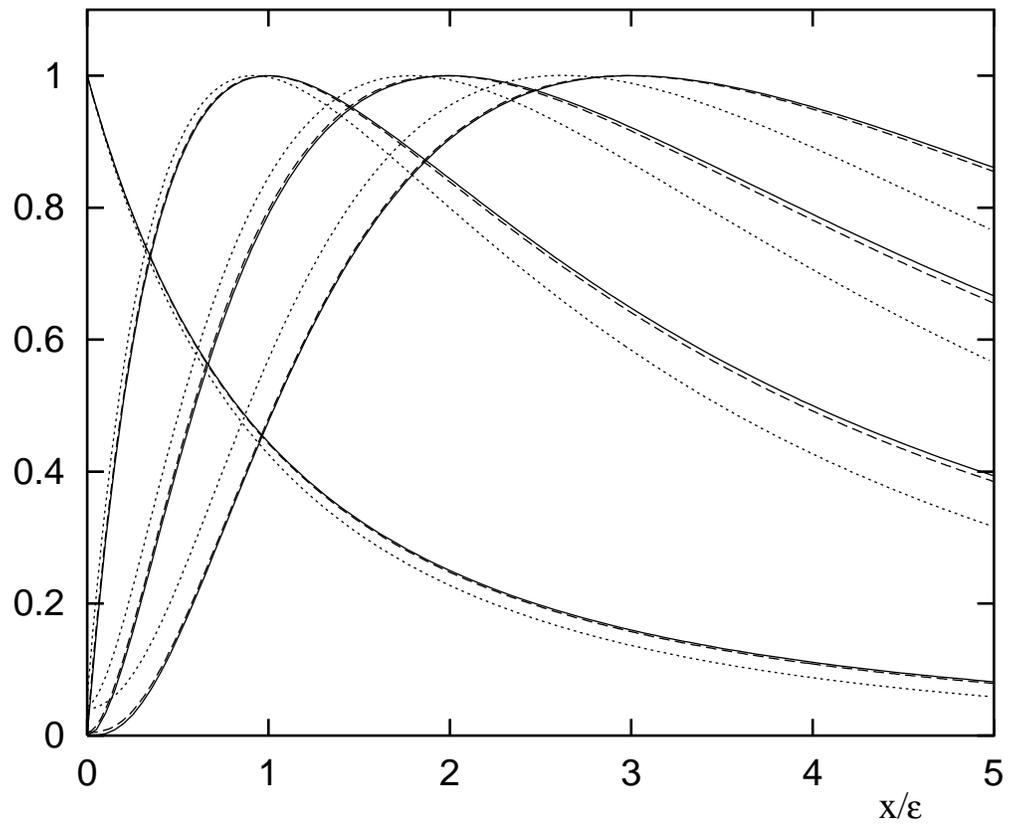
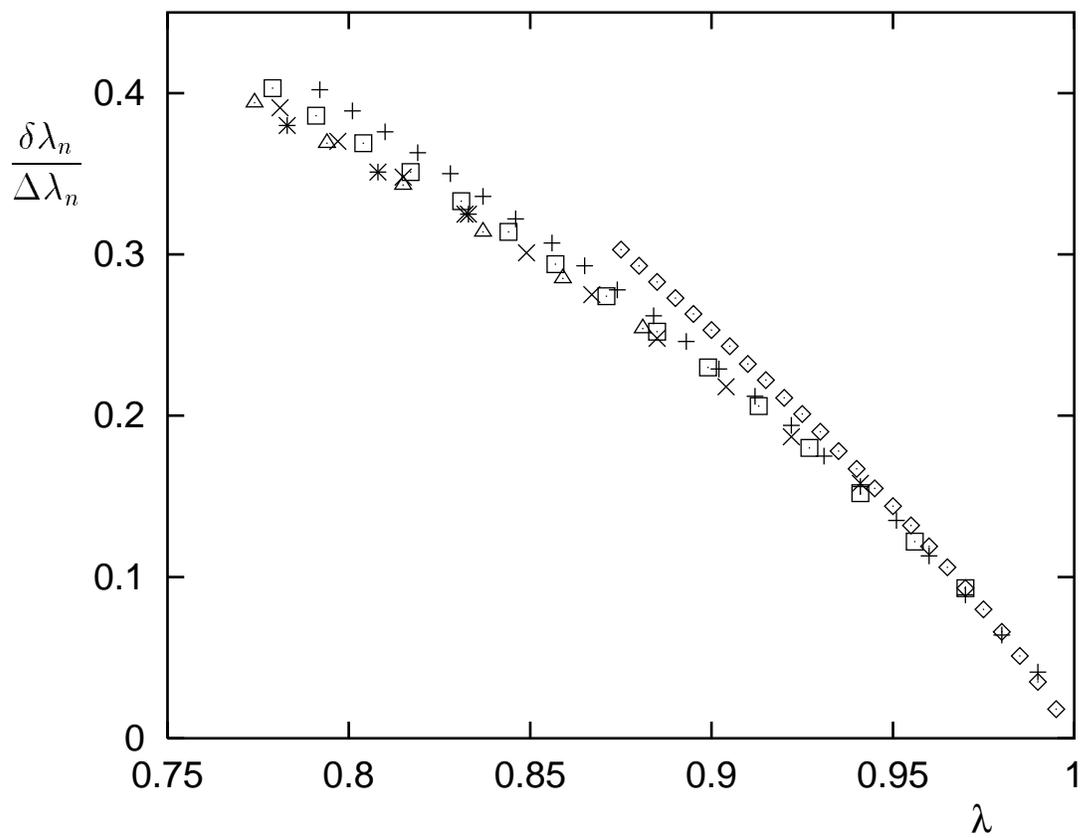
\end{document}